\documentclass[twoside,12pt]{article}
\usepackage{amssymb}
\date{}
\begin{document}
\baselineskip=7truemm
\setcounter{section}{0}
\title{ \bf Exact Propagator of a two dimensional anisotropic Harmonic Oscillator in the presence of a   Magnetic Field }
\author{{\bf Jose M. Cerver\'o} \\ {\small \bf Departamento de F\'{\i}sica Fundamental}. {\small \bf Facultad de Ciencias.} \\
{\small \bf Universidad de Salamanca.} {\small \bf 37008 Salamanca. Spain}}
\maketitle
\begin{abstract}
\noindent  In this paper we solve exactly the problem of the spectrum and Feynman propagator of a charged particle submitted to both an anharmonic oscillator in the plane and a constant and homogeneous magnetic field of arbitrary strength aligned with the perpendicular direction to the plane. As we shall see in the beginning of the letter the Hamiltonian, being a quadratic form,  is easily diagonalizable and the Classical Action can be used to construct the exact Feynman Propagator using the Stationary Phase Approximation. The result is useful for the treatment of quasi two dimensional samples in the field of  magnetic effects in nano-structures and quantum optics. The presented solution, after minor extensions, can also be used for motion in three dimensions, and in fact it has been used for years in such cases. Also it can be used as a good exercise of a Feynman Path Integral that can be calculated easily with just the help of the Classical Action.
\end{abstract}
\vskip 0.4cm
$\quad${\bf PACS Number: ${\bf 03.65.} -$w}
\vskip 0.3 true in
\section{Introduction.}
\noindent The problem of quantum charged particles in the presence of electric and magnetic fields has always remained on target of any theoretical physicist who wish  to predict  the behavior of one or a swarm of such particles imbedded in the structure of a solid. The {\bf exact solution} in three dimensions has eluded these attempts for years. However, when the particles remain in a {\bf two dimensional solid} the problem of the {\bf exact solution} becomes both feasible and possible.  Experimentally, the last situation has only been possible to be realized until very recently in the laboratory  thanks to the technological means that allow the experimentalists to produce quasi dimensional samples of metals and semimetals. One can induce large magnetic behavior combining them with ferromagnetic 2D materials such as ${\bf La_{0,7}Sr_{0,3}MnO_3/SrTiO_3}$ and ${\bf K_2CuF_4}$. Indeed we should also mention the recent discovery of {\bf Magnetic Graphene} which falls in the class of these two dimensional samples when combined with the aforementioned substrates. The propagator can also be used in three dimensions with some little arrangements. However I restrict myself to the two dimensional case given the importance of two dimensional solids nowadays.

However the stationary solution of a bounded Hamiltonian is only of academic interest as we would like to know the evolution of the particle inside of the two dimensional solid. This goal can only be achieved if we know the propagator of the particle(s) along the time and the properties that may or may not change in the sample: phase transitions from metal to semimetal or to an insulator phase, the behavior of the Anderson localization, the magneto-optics properties of the solid and the like. This wealth of physical effects can be studied if we know the Feynman propagator of the system and the help of computing algebraic calculus. The aim of this paper is to show that these results can be attained with little effort and a touch of elegance.

The problem of the exact solution of the Feynman propagator of the system simplifies a great deal with the help of the Stationary Phase Approximation that {\bf turns to be exact} in the case of a quadratic Hamiltonians. In this paper we use the most general two-dimensional Hamiltonian with magnetic effects provided by a magnetic field pointing in the perpendicular direction of the plane. The rest of the interaction is represented by anisotropic harmonic oscillators lying in the plane along perpendicular directions. 

I would like to finish this Introduction with a set of References. The Feynman Path Integrals \cite{FEYNMAN} are well known from a historical perspective \cite{FEYNMANT} to an encyclopedic account with many examples \cite{KLEINERT}. Also this Quantum Mechanical problem has been partially treated many years ago in references \cite{KOKICAS} and \cite{DAVIES}.
\section{Quantum Mechanical Hamiltonian}
  
The Hamiltonian described in the Introduction has the form: 
\begin{eqnarray}
\hat H={1\over 2m}\{\hat p_x - {e\over c}\hat A_x\}^2 + {1\over 2m}\{\hat p_y - {e\over c}\hat A_y\}^2
+ {1\over 2}m\omega_1^2\hat x^2   + {1\over 2}m\omega_2^2\hat y^2
\end{eqnarray}
One can select the following gauge for the magnetic vector that produces a constant magnetic field of arbitrary strength and perpendicular to the plane:
\begin{eqnarray}
A_x = - \{{\omega_2\over \omega_1 + \omega_2}\}B_0 y \quad;\quad 
A_y = +   \{{\omega_1\over \omega_1 + \omega_2}\}B_0 x  \quad;\quad A_3 = 0
\end{eqnarray}
and the Hamiltonian reads:
\begin{eqnarray}
\hat H={\hat p_x^2\over 2m}  + {\hat p_y^2\over 2m} + {1\over 2}m\omega_1^2 [1 +  {\omega_0^2\over (\omega_1 +
\omega_2)^2}]\hat x^2 + \nonumber \\ 
+ {1\over 2}m\omega_2^2 [1 +  {\omega_0^2\over (\omega_1 + \omega_2)^2}]\hat y^2  
+ {\omega_0\over \omega_1 + \omega_2}(\omega_2 \hat y \hat p_x - \omega_1 \hat x  \hat p_y)
\end{eqnarray}
with
\begin{eqnarray}
\omega_0 = {eB_0 \over mc}
\end{eqnarray}
Let us now call:
\begin{eqnarray}
\gamma = [1 + {\omega_0^2\over (\omega_1 + \omega_2)^2}]^{1\over 2}
\end{eqnarray}
and we scale the position and momentum operators in a scale free way in the form:
\begin{eqnarray}
\hat X = ({m \gamma\omega_1\over \hbar})^{1\over 2} \hat x \qquad\qquad\qquad\qquad\quad
\hat Y = ({m \gamma\omega_2\over \hbar})^{1\over 2} \hat y \quad 
\end{eqnarray}
\begin{eqnarray}
\hat P_x = ({\hbar m \gamma\omega_1})^{-{1\over 2}} \hat p_x\qquad\qquad\qquad\qquad
\hat P_y = ({\hbar m \gamma\omega_2})^{-{1\over 2}} \hat p_y  
\end{eqnarray}
Again, with this change of variables the Hamiltonian takes the form:
\begin{eqnarray}
\{{2\over \hbar\omega_1\gamma}\}\hat H= (\hat P_x^2 + \hat X ^2)  +  
({\omega_2\over\omega_1}) (\hat P_y^2 + \hat Y^2) +
{2\omega_0\over [(\omega_1 + \omega_2)^2 +\omega_0^2]^{1\over 2}} ({\omega_2\over\omega_1})^{1\over 2}
 (\hat Y \hat P_x - \hat X \hat P_y)   
\end{eqnarray}
that looks as a quadratic form that we can easily diagonalize:
\[ {\bf \epsilon} (\lambda) = \left| \begin{array}{cccc}
1 -\lambda& 0 & 0 & {c\over 2}\\ 0&1 -  \lambda &  -{c\over 2} & 0\\ 0 & - {c\over 2} &b- \lambda&0\\
 {c\over 2} &0 &0 & b-\lambda  \end{array} \right| \]    \qquad\qquad 
\noindent where $b$ and $c$ are constants:
\begin{eqnarray}
b= ({\omega_2\over\omega_1}) \quad;\quad 
c= {2\omega_0 b^{1\over 2}\over [\omega_0^2 + (\omega_1 + \omega_2)^2 ]^{1\over 2}}
\quad;\quad \gamma = {[\omega_0^2 + (\omega_1 + \omega_2)^2 ]^{1\over 2} \over (\omega_1 + \omega_2)}
\end{eqnarray}
The non vanishing real roots of the quartic equation take the form:
\begin{eqnarray}
\epsilon_{\pm}={1\over 2 \omega_1}\{{(\omega_1 + \omega_2)} \pm \sqrt{{(\omega_1 - \omega_2)^2} + {4\omega_0^2
\omega_1 \omega_2\over [\omega_0^2 + (\omega_1 + \omega_2)^2 ]}}\} = \nonumber \\
= {(\omega_1 + \omega_2)\over 2 \omega_1}
\{1 \pm \sqrt{[\omega_0^2 + (\omega_1 - \omega_2)^2]\over [\omega_0^2 + (\omega_1 + \omega_2)^2]}\} 
\qquad\qquad
\end{eqnarray}
and multiplying the Hamiltonian by  ${\hbar\over 2}\{\omega_1\gamma\}$,  one finally obtains the following {\bf eigenvalues}:
\begin{eqnarray}
E_{\pm} = {\hbar \over 4} \{[\omega_0^2 + (\omega_1 + \omega_2)^2 ]^{1\over 2} \pm 
[\omega_0^2 + (\omega_1 - \omega_2)^2 ]^{1\over 2}\} =  
{\hbar\over 4}\{\Omega_+ \pm \Omega_-\}
\end{eqnarray}
or in a more tractable manner:
\begin{eqnarray}
\epsilon_{n_1, n_2}={\hbar\over 4}\{(\Omega_+ +\Omega_-)(n_1 + {1\over 2}) +
(\Omega_+  - \Omega_-)(n_2 + {1\over 2})\} = \nonumber \\  = {\hbar\over 4}\{\Omega_+ (n_1 + n_2 +1) +
\Omega_- (n_1 - n_2)\}  \qquad\qquad 
\end{eqnarray}
If one diagonalizes the isotropic harmonic oscillator in both cartesian and angular coordinates the Landau eigenlevels are labeled respectively by the quantum numbers $(n_1 , n_2)$ and $(n_r, m)$. Both eigenlevels take the form:
\begin{eqnarray}
(n_1 + n_2 +1) = (2n_r + \mid m \mid + 1)   \quad\quad where \quad\quad m = (n_1 - n_2 ) 
\end{eqnarray}
\noindent Therefore the Hamiltonian exhibits a complete breaking of the magnetic degeneracy. This effect has nothing to do with the spin-orbit coupling as the spin degrees of freedom are absent from the Hamiltonian. It is in turn closely related to the Landau orbital coupling interacting with the magnetic field. The non trivial question of writing down the wave functions has been largely discussed in \cite{SEBAWE} and \cite{CEDERBAUM}. The interested reader is invited to address himself to these  references. 
\vskip 0.5cm
\section{Lagrangian Formalism and construction of the Propagator}
The Lagrangian of the system has the obvious form:
\begin{eqnarray}
{\cal L}(x_i, v_i) =  {1\over 2}m \{{\dot x}^2 + {\dot y}^2 \} + 
{e\over c}\{\dot x  A_x + \dot y A_y\} - {1\over 2}m\omega_1^2  x^2   - {1\over 2}m\omega_2^2 y^2 
\end{eqnarray}
and now select a {\bf different gauge},  but representing also a constant magnetic field in the perpendicular 
direction to the plane:
\begin{eqnarray}
\bar A_x = - {B_0\over 2} y  \quad;\quad 
\bar A_y = + {B_0\over 2} x  \quad;\quad \bar A_3 = 0
\end{eqnarray}
The change of gauge causes no problems in the final solution of the propagator as it was pointed out many years ago by Bialynicki-Birula \cite{BB}. The propagator and quantum amplitudes in general are invariant under a gauge transformation. In our case:
\begin{eqnarray}
\bar A_x =  - \{{\omega_2\over \omega_1 + \omega_2}\}B_0 y - \partial_x {\bf \Lambda} \quad;\quad  \bar A_y = +   \{{\omega_1\over \omega_1 + \omega_2}\}B_0 x- \partial_y {\bf \Lambda} 
\end{eqnarray}
It is easy to find out the actual fom of  ${\bf \Lambda}(x , y, z)$, obviously   independent of $z$. One finds:
\begin{eqnarray}
{\bf \Lambda}(x , y)  =   {B_0\over 2} {(\omega_1 - \omega_2) \over (\omega_1 + \omega_2)}xy
\end{eqnarray}
which vanishes for the isotropic limit. For a throughout and recent discussion on gauge independence of the propagator see reference \cite{BANDRAUK}.
\newpage
\noindent The Lagrangian looks now as:
\begin{eqnarray}
{\cal L}(x, y, {\dot x}, {\dot y}) =  {m\over 2} \{{\dot x}^2 + {\dot y}^2    
- (\omega_1^2  x^2 + \omega_2^2 y^2) + \omega_0 (x \dot y - y \dot x) \}  
\end{eqnarray}
and the Equations of Motion are:
\begin{eqnarray}
\ddot x  + \omega_1^2 x  = \omega_0 \dot y  \qquad;\qquad  \ddot y + \omega_2^2 y = - \omega_0 \dot x 
\end{eqnarray}
The {\bf real solutions} of these Equations are always harmonic functions that we can write of the form:
\begin{eqnarray}
x(t) = {\bf A} \cos{1\over 2}(\Omega_+ + \Omega_-)t + {\bf B} \sin{1\over 2}(\Omega_+ + \Omega_-)t + \nonumber \\
+ {\bf C} \cos{1\over 2}(\Omega_+ - \Omega_-)t + {\bf D} \sin{1\over 2}(\Omega_+ - \Omega_-)t  \\
y(t) = -{\bf \Lambda_1}\{{\bf A} \sin{1\over 2}(\Omega_+ + \Omega_-)t - {\bf B} \cos{1\over 2}(\Omega_+ + \Omega_-)t \}   -\nonumber \\- {\bf \Lambda_2}\{{\bf C} \sin{1\over 2}(\Omega_+ - \Omega_-)t - {\bf D} \cos{1\over 2}(\Omega_+ - \Omega_-)t\}   
\end{eqnarray}  
where: 
\begin{eqnarray}
{\bf \Lambda_1} = {(\Omega_+ + \Omega_-)^2 - 4\omega_1^2 \over 2\omega_0 (\Omega_+ + \Omega_-)} \qquad;\qquad
{\bf \Lambda_2} = {(\Omega_+ - \Omega_-)^2 - 4\omega_1^2 \over 2\omega_0 (\Omega_+ - \Omega_-)}
\end{eqnarray} 
We shall name: $x(t=0) = x_1$ , $y(t=0) = y_1$ to the points at time equal zero and $x(t=T) = x_2$ , $y(t=T) = y_2$,
those points at an arbitrary evolution time $T$. Thus, one can see trivially that the following relationships hold:
\begin{eqnarray}
x_1 = {\bf A} + {\bf C} \qquad;\qquad y_1 = {\bf \Lambda_1} {\bf B} + {\bf \Lambda_2} {\bf D}
\end{eqnarray}
\begin{eqnarray}
x_2 = {\bf A} \cos{1\over 2}(\Omega_+ + \Omega_-)T + {\bf B} \sin{1\over 2}(\Omega_+ + \Omega_-)T + \nonumber \\
+ {\bf C} \cos{1\over 2}(\Omega_+ - \Omega_-)T + {\bf D} \sin{1\over 2}(\Omega_+ - \Omega_-)T
\end{eqnarray}
\begin{eqnarray}
y_2 = -{\bf \Lambda_1}\{{\bf A} \sin{1\over 2}(\Omega_+ + \Omega_-)T - {\bf B} \cos{1\over 2}(\Omega_+ + \Omega_-)T \} - \nonumber \\
- {\bf \Lambda_2}\{{\bf C} \sin{1\over 2}(\Omega_+ - \Omega_-)T - {\bf D} \cos{1\over 2}(\Omega_+ - \Omega_-)T\} 
\end{eqnarray}
Next we shall be concerned with the evaluation of the classical action of the anisotropic net in the presence of a uniform and constant magnetic field. The various steps of the calculation are listed in the {\bf APPENDIX} at the end of the paper. For the moment  we shall be concerned formally with the action derived from this Lagrangian ${\bf S_{cl}^{AMN}}$ having the form:
\begin{eqnarray}
{\bf S_{cl}^{AMN}} [x(T), y(T)] &=& \int_{t_1}^{t_2}{\cal L}(x, y, {\dot x}, {\dot y})dt   = \nonumber \\    
&=&  {m\over 2}  \int_{t_1}^{t_2}dt\{{\dot x}^2 + {\dot y}^2 - (\omega_1^2  x^2 + \omega_2^2 y^2) + \omega_0 (x \dot y - y \dot x) \} \qquad\qquad 
\end{eqnarray}
where we have used (16). Next we integrate by parts the terms ${\dot x}^2$ and ${\dot y}^2$ leading easily to:
\begin{eqnarray}
{\bf S_{cl}^{AMN}} [x(T), y(T)]  = {m\over 2}\{x \dot x  +  y \dot y   \} \mid_{t_1}^{t_2} -\nonumber \\ 
- {m\over 2} \int_{t_1}^{t_2}dt \{\{\ddot x  + \omega_1^2 x  - \omega_0 \dot y\}x +\{ \ddot y + \omega_2^2 y + \omega_0 \dot x\}y \}
\end{eqnarray} 
The integral of the right hand side is trivially zero as it contains only the equations of motion (17). This property has been used in other contexts in reference \cite{MECHANICS} for the general case of kinetic energy plus an arbitrary potential depending just upon the coordinates. One finally founds:
\begin{eqnarray}
{\bf S_{cl}^{AMN}} [x(T), y(T)] = {m\over 2}\{x_2\dot x_2  - x_1\dot x_1 +  y_2\dot y_2 - y_1\dot y_1\} 
\end{eqnarray} 
and substituting the values of the coordinates and their derivatives one finally obtains the following expression for the action ${\bf S^{AMN}}_{cl}[x(T), y(T)] $:
\begin{eqnarray}
{\bf S_{cl}^{AMN}} [x(T), y(T)] = {m\over 2{\bf D}(T)}\{{\bf a_1}(T)(x^2_1 + x^2_2) + {\bf a_2}(T)(y^2_1 + y^2_2) - 2
{\bf b_1}(T) x_1x_2 - \nonumber \\
 - 2{\bf b_2}(T) y_1y_2 +  4 {\bf c_1}(T) (x_1y_2 - x_2y_1) + {\bf c_2}(T) (x_2y_2 - x_1y_1)\} \qquad\qquad
\end{eqnarray} 
The {\bf Amplitude of the Propagator} is now calculated with the help of the {\bf VanVleck-DeWitt-Morette}  Determinant. As it is well known the form of the Determinant is:
\begin{eqnarray}
{\bf A}^{\bf AMN}(T) = \{{1\over {2 \pi i\hbar}}\}^{d\over 2}\mid {\bf Det} 
{\bf M}\mid^{1\over 2} = \{{1\over {2 \pi i\hbar}}\}^{d\over 2}\mid {\bf Det}{{\partial^2 {\bf S}^{\bf AMN}_{cl}\over \partial x_i \partial y_j}} \mid^{1\over 2}
\end{eqnarray} 
where $d$ is the number of {\bf spatial} dimensions of the problem. In our case $d = 2$. Furthermore the form of the Determinant is:
\[ {\bf M} = \left| \begin{array}{cc}
{\partial^2 {\bf S}^{\bf AMN}_{cl}/\partial x_1 \partial x_2}&{\partial^2 {\bf S}^{\bf AMN}_{cl}/\partial y_1 \partial x_2} 
\\{\partial^2 {\bf S}^{\bf AMN}_{cl}/\partial x_1 \partial y_2}&{\partial^2 {\bf S}^{\bf AMN}_{cl}/\partial y_1 \partial y_2}   \end{array} \right|.\]  
 The result in our case leads to:
\begin{eqnarray}
{\bf A}^{\bf AMN}(T) = { \mid {\bf Det} {\bf M}\mid^{1\over 2}\over {2 \pi i\hbar}} &=&  {m\over 2\pi i \hbar
{\bf D}(T)}  \sqrt{{\bf b_1}(T){\bf b_2}(T) + 4{\bf c_1}^2(T)} = \nonumber \\
&=& {m\over {2 \pi i\hbar}} \sqrt{{(\omega_1 \omega_2 )(\Omega_+ \Omega_-) \over  {\bf D}(T)}}
\end{eqnarray} 
Substituting ${\bf D}(T)$  we  obtain (with $T \Longrightarrow t$):
\begin{eqnarray}
{\bf A}^{\bf AMN}(t) =  {m  \Omega_+ \Omega_-  \over {2 \pi i\hbar}} \left [ { \omega_1 \omega_2  
\over  (\omega_1 + \omega_2)^2 \Omega_-^2 \sin^2{\Omega_+ t \over 2} -
(\omega_1 - \omega_2)^2 \Omega_+^2 \sin^2{\Omega_-  t \over 2}} \right ]^{1\over 2}
\end{eqnarray} 
and the propagator takes the final form:
\begin{eqnarray}
{\bf G}^{\bf AMN}(x_1, y_1, x_2, y_2, t ) = {\bf A}^{\bf AMN}(t) \exp\{{i\over \hbar}
{\bf S_{cl}^{AMN}} [x_1, y_1, x_2, y_2, t] \}
\end{eqnarray} 
where
\begin{eqnarray}
{\bf S_{cl}^{AMN}} [x_1, y_1, x_2, y_2, t]  = {m\over 2{\bf D}(t)}\{{\bf a_1}(t)(x^2_1 + x^2_2) + {\bf a_2}(t)(y^2_1 + y^2_2) - 2{\bf b_1}(t) x_1x_2 - \nonumber \\
- 2{\bf b_2}(t) y_1y_2 +  2 {\bf c_1}(t) (x_1y_2 - x_2y_1) + {\bf c_2}(t) (x_2y_2 - x_1y_1)\} \qquad\qquad
\end{eqnarray} 
The expressions for  ${\bf a_1}(t), {\bf a_2}(t), {\bf b_1}(t) , {\bf b_2}(t) , {\bf c_1}(t)$ and ${\bf c_2}(t)$  can be found in the {\bf APPENDIX}  but now we have substituted everywhere $T \Longrightarrow t$ as in the ${\bf D}(T)$ expression.
 As a final word one can say that these seemingly tedious calculations simplify greatly when one uses any of the available packages of algebraic calculations such as ${\it MATHEMATICA}^T$ or ${\it MAPLE}^T$. Also limits in the case of  vanishing magnetic fields or isotropic harmonic oscillator (i.e. $\omega_1 = \omega_2$) have been discussed in \cite{URRUTIA}. One of the main goals of this work has been precisely to put at work these software applications to the non trivial field of {\bf path integrals} and its use in material science and quantum optics: for instance propagating a gaussian wave packet in two dimensional magnetic solids when the gaussian represents charged particles or intense femtosecond light pulses. A particular case is the one of a "Schr\"odinger cat" propagating in the net. The wave function can be simulated as two widely separated gaussians charged with  "$-2e$-charge"  as in a Cooper pair \cite{OCONNELL}.
 \begin{eqnarray}
 \Psi = {1\over (8\pi \sigma^2)^{1/4}}\otimes {1\over (1+ e^{-{a_0^2/ 8 \sigma^2}})^{1/2}}\{ \exp\{- {(x+{a_0\over 2})^2\over 4\sigma^2}\} + \exp\{- {(x- {a_0\over 2})^2\over 4\sigma^2}\}
\end{eqnarray} 
where $a_0 = {\hbar^2 \over me^2}$ in atomic units containing the electric charge and $\sigma^2$ is the variance of each packet. Although these projects for Master Thesis or Graduate Courses can be considered as a little cumbersome, they might attract the interest of the young physicists to the condensed matter field in which much of the research  is being done nowadays.
\newpage
\noindent $\bullet$ {\bf APPENDIX.} {\bf Calculations and Useful definitions  of interest}
\vskip 0.1cm
\noindent To obtain $\{{\bf A}, {\bf B}, {\bf C} , {\bf D}\}$ in equations (23)-(25) as functions of  $\{x_1, y_1, x_2 , y_2\}$.    I have used for the calculation of the inverse $4 \otimes 4$ matrix with the help of   the ${\it MATHEMATICA}^T$ package:
\begin{eqnarray}
\left(\matrix{{\bf A}\cr {\bf B}\cr {\bf C}\cr {\bf D}}\right) =  \left(\matrix{u_{11}&u_{12}&u_{13}
&u_{14}\cr u_{21}&u_{22}&u_{23}&u_{24}\cr u_{31}&u_{32}&u_{33}&u_{34}\cr u_{41}&u_{42}&u_{43}&u_{44}}\right) 
\left(\matrix{x_1\cr y_1\cr x_2 \cr y_2}\right) 
\end{eqnarray}
where the matrix elements and the determinant  ${\bf \Delta}$  of (24) are listed below:
\begin{eqnarray}
a_{11} &=& {\bf \Lambda_2} \{ {\bf \Lambda_1} \cos{1\over 2}(\Omega_+ + \Omega_-)T\cos{1\over 2}(\Omega_+ - \Omega_-)T   +  {\bf \Lambda_2}\sin{1\over 2}(\Omega_+ + \Omega_-)T\sin{1\over 2}(\Omega_+ - \Omega_-)T 
 -{\bf \Lambda_1}\}  \nonumber \\
a_{12} &=&  {\bf \Lambda_1} \cos{1\over 2}(\Omega_+ + \Omega_-)T \sin{1\over 2}(\Omega_+ - \Omega_-)T - 
{\bf \Lambda_2} \sin{1\over 2}(\Omega_+ + \Omega_-)T \cos{1\over 2}(\Omega_+ - \Omega_-)T  \nonumber \\
a_{13} &=& {1\over 2} {\bf \Lambda_1} {\bf \Lambda_2} \sin{1\over 2}(\Omega_+ + \Omega_-)T\sin{1\over 2}(\Omega_+ - \Omega_-)T  \nonumber \\
a_{14} &=&  {\bf \Lambda_2} \sin{1\over 2}(\Omega_+ + \Omega_-)T - {\bf \Lambda_1}\sin{1\over 2}(\Omega_+ - \Omega_-)T  \nonumber \\
a_{21} &=&   {\bf \Lambda_2}\{{\bf \Lambda_1} \sin{1\over 2}(\Omega_+ + \Omega_-)T \cos{1\over 2}(\Omega_+ - \Omega_-)T- {\bf \Lambda_2} \cos{1\over 2}(\Omega_+ + \Omega_-)T \sin{1\over 2}(\Omega_+ - \Omega_-)T \}
 \nonumber \\
a_{22} &=&  {\bf \Lambda_1} \sin{1\over 2}(\Omega_+ + \Omega_-)T \sin{1\over 2}(\Omega_+ - \Omega_-)T   +
{\bf \Lambda_2} \cos{1\over 2}(\Omega_+ + \Omega_-)T \cos{1\over 2}(\Omega_+ - \Omega_-)T - {\bf \Lambda_2}
 \nonumber \\
a_{23} &=& {\bf \Lambda_2}\{ - {\bf \Lambda_1} \sin{1\over 2}(\Omega_+ + \Omega_-)T  +  {\bf \Lambda_2} \sin{1\over 2}(\Omega_+ - \Omega_-)T\}  \nonumber \\
a_{24} &=& {1\over 2}  {\bf \Lambda_2} \sin{1\over 2}(\Omega_+ + \Omega_-)T\sin{1\over 2}(\Omega_+ - \Omega_-)T
\nonumber \\
a_{31} &=& {\bf \Lambda_1} \{ {\bf \Lambda_2} \cos{1\over 2}(\Omega_+ + \Omega_-)T\cos{1\over 2}(\Omega_+ - \Omega_-)T   +  {\bf \Lambda_1}\sin{1\over 2}(\Omega_+ + \Omega_-)T\sin{1\over 2}(\Omega_+ - \Omega_-)T  - {\bf \Lambda_2}\}  \nonumber \\
a_{32} &=&  - {\bf \Lambda_1} \cos{1\over 2}(\Omega_+ + \Omega_-)T \sin{1\over 2}(\Omega_+ - \Omega_-)T +
{\bf \Lambda_2} \sin{1\over 2}(\Omega_+ + \Omega_-)T \cos{1\over 2}(\Omega_+ - \Omega_-)T   \nonumber \\
a_{33} &=& -{1\over 2} {\bf \Lambda_1} {\bf \Lambda_2} \sin{1\over 2}(\Omega_+ + \Omega_-)T\sin{1\over 2}(\Omega_+ - \Omega_-)T   \nonumber \\
a_{34} &=&  - {\bf \Lambda_2} \sin{1\over 2}(\Omega_+ + \Omega_-)T + {\bf \Lambda_1}\sin{1\over 2}(\Omega_+ - \Omega_-)T    \nonumber \\
a_{41} &=&   {\bf \Lambda_1}\{- {\bf \Lambda_1} \sin{1\over 2}(\Omega_+ + \Omega_-)T \cos{1\over 2}(\Omega_+ - \Omega_-)T + {\bf \Lambda_2} \cos{1\over 2}(\Omega_+ + \Omega_-)T \sin{1\over 2}(\Omega_+ - \Omega_-)T \}
 \nonumber \\
a_{42} &=&  {\bf \Lambda_2} \sin{1\over 2}(\Omega_+ + \Omega_-)T \sin{1\over 2}(\Omega_+ - \Omega_-)T   +
{\bf \Lambda_1} \cos{1\over 2}(\Omega_+ + \Omega_-)T \cos{1\over 2}(\Omega_+ - \Omega_-)T - {\bf \Lambda_1}
 \nonumber \\
a_{43} &=& {\bf \Lambda_1}\{ {\bf \Lambda_1} \sin{1\over 2}(\Omega_+ + \Omega_-)T  -  {\bf \Lambda_2} \sin{1\over 2}(\Omega_+ - \Omega_-)T\}   \nonumber \\
a_{44} &=&  - {1\over 2}  {\bf \Lambda_1} \sin{1\over 2}(\Omega_+ + \Omega_-)T\sin{1\over 2}(\Omega_+ - \Omega_-)T
\nonumber
\end{eqnarray}
\vskip 0.2cm
\noindent and the determinant takes the form:
\begin{eqnarray}
{\bf \Delta}Ê= \{({\bf \Lambda_1} - {\bf \Lambda_2})^2  \sin{1\over 2}(\Omega_+ + \Omega_-)T \sin{1\over 2}(\Omega_+ -\Omega_-)T - 2 {\bf \Lambda_1} {\bf \Lambda_2} (1 - \cos \Omega_- T  ) \qquad
\end{eqnarray}
Likewise one can easily calculate $\{\dot x_1, \dot y_1, \dot x_2 , \dot y_2\}$
\begin{eqnarray}
\dot x_1 &=& {1\over 2}(\Omega_+ + \Omega_-){\bf B} + {1\over 2}(\Omega_+ - \Omega_-){\bf D} \\
\dot y_1 &=&  -{1\over 2}(\Omega_+ + \Omega_-){\bf \Lambda_1} {\bf A} -{1\over 2}(\Omega_+ - \Omega_-){\bf \Lambda_2} 
{\bf C}  
\end{eqnarray}
\begin{eqnarray}
\dot x_2 = -{1\over 2}(\Omega_+ + \Omega_-)\{{\bf A} \sin{1\over 2}(\Omega_+ + \Omega_-)T - 
{\bf B} \cos{1\over 2}(\Omega_+ + \Omega_-)T\} - \nonumber \\
- {1\over 2}(\Omega_+ - \Omega_-) \{{\bf C} \sin{1\over 2}(\Omega_+ - \Omega_-)T - {\bf D}
\cos{1\over 2}(\Omega_+ - \Omega_-)T\}  
\end{eqnarray}
\begin{eqnarray}
\dot y_2 = -{1\over 2}(\Omega_+ + \Omega_-){\bf \Lambda_1}\{{\bf A} \cos{1\over 2}(\Omega_+ + \Omega_-)T + {\bf B} \sin{1\over 2}(\Omega_+ + \Omega_-)T\} - \nonumber \\
 -{1\over 2}(\Omega_+ - \Omega_-){\bf \Lambda_2}\{{\bf C} \cos{1\over 2}(\Omega_+
- \Omega_-)T + {\bf D} \sin{1\over 2}(\Omega_+ - \Omega_-)T\} 
\end{eqnarray}

The derivatives then take the form: 
\begin{eqnarray}
\dot x_1 = {1\over {\bf D}(t) }\{- {\bf a_1}(T) x_1 +  {\bf b_1}(T) x_2 -  2{\bf c_1}(T)y_2 +  {\bf f_1}(T)y_1\} 
\end{eqnarray}
\begin{eqnarray}
\dot x_2 = {1\over {\bf D}(t) }\{- {\bf b_1}(T) x_1 +  {\bf a_1}(T) x_2 -  2{\bf c_1}(T)y_1 +  {\bf f_1}(T)y_2\}
\end{eqnarray}
\begin{eqnarray}
 \dot y_1 ={1\over {\bf D}(t) }\{ - {\bf a_2}(T) y_1 +  {\bf b_2}(T) y_2 + 2{\bf c_1}(T)x_2 +  {\bf f_2}(T)x_1\}
\end{eqnarray}
\begin{eqnarray}
 \dot y_2 = {1\over {\bf D}(t) }\{ - {\bf b_2}(T) y_1 +  {\bf a_2}(T) y_2 + 2{\bf c_1}(T)x_1 +  {\bf f_2}(T)x_2 \}
\end{eqnarray}
where:
\begin{eqnarray}
{\bf a_1}(T) = {\omega_1 \over 2}\{\Omega_-(\omega_1 +
\omega_2)\sin{\Omega_+ T}  - \Omega_+(\omega_1 - \omega_2)\sin{\Omega_- T}\}
\end{eqnarray}
\begin{eqnarray}
{\bf a_2}(T) = {\omega_2 \over 2}\{\Omega_- (\omega_1 + \omega_2)\sin{\Omega_+ T} + \Omega_+(\omega_1 - \omega_2)\sin{\Omega_- T} \}
\end{eqnarray}
\begin{eqnarray}
{\bf b_1}(T) = \omega_1\{\Omega_- (\omega_1 + \omega_2) \sin{\Omega_+T\over 2} 
\cos{\Omega_- T\over 2} - \Omega_+(\omega_1 -\omega_2) 
\cos{\Omega_+T\over 2}\sin{\Omega_-T \over 2}\} \quad
\end{eqnarray}
\begin{eqnarray}
{\bf b_2}(T) = \omega_2 \{\Omega_- (\omega_1 + \omega_2) \sin{\Omega_+T \over 2} 
\cos{\Omega_- T\over 2} + \Omega_+(\omega_1 -\omega_2)\cos{\Omega_+T\over 2}
\sin{\Omega_-T \over 2}\} \quad
\end{eqnarray}
\begin{eqnarray}
{\bf c_1}(T) =  \omega_0 \omega_1 \omega_2 \{\sin{\Omega_+T \over 2} \sin{\Omega_-T \over 2}\}  
\end{eqnarray}
\begin{eqnarray}
{\bf c_2}(T) = \omega_0(\omega_1 + \omega_2)(\omega_1 - \omega_2)
\{({\Omega_+\over \Omega_-})\sin^2{\Omega_- T\over 2} - ({\Omega_-\over \Omega_+}) 
\sin^2{\Omega_+T\over 2}\} 
\end{eqnarray}
\begin{eqnarray}
{\bf f_1}(T) =  \omega_0\{\omega_2(\omega_1 -\omega_2) ({\Omega_+\over \Omega_-}) 
\sin^2{\Omega_- T \over 2} + \omega_2(\omega_1 +\omega_2) ({\Omega_-\over \Omega_+}) 
\sin^2{\Omega_+ T\over 2}\}
\end{eqnarray}
\begin{eqnarray}
{\bf f_2}(T) =   \omega_0\{\omega_1(\omega_1 -\omega_2) ({\Omega_+\over \Omega_-}) 
\sin^2{\Omega_- T\over 2} - \omega_1(\omega_1 +\omega_2) ({\Omega_-\over \Omega_+}) 
\sin^2{\Omega_+ T\over 2}\}
\end{eqnarray}
\begin{eqnarray}
{\bf D}(T) = (\omega_1 + \omega_2)^2 ({\Omega_-\over \Omega_+}) \sin^2{\Omega_+T \over 2} -
(\omega_1 - \omega_2)^2 ({\Omega_+\over \Omega_-}) \sin^2{\Omega_- T \over 2}
\end{eqnarray}
Notice the interesting relationship:
\begin{eqnarray}
{\bf c_2}(T) = {\bf f_1}(T) + {\bf f_2}(T)
\end{eqnarray}  
\vskip 0.2cm

\noindent The following constants have been defined in the body of the Paper as:
\begin{eqnarray}
\Omega_+ =  \sqrt{\omega_0^2 + (\omega_1 + \omega_2)^2}  \qquad&;&\qquad
\Omega_- =  \sqrt{\omega_0^2 + (\omega_1 - \omega_2)^2}  \\
\Omega_1 =  {1\over 2}(\Omega_+  + \Omega_-)  \qquad&;&\qquad
\Omega_2 =   {1\over 2}(\Omega_+  - \Omega_-) 
\end{eqnarray}
\begin{eqnarray}
{\bf \Lambda_1} = {(\Omega_+ + \Omega_-)^2 - 4\omega_1^2 \over 2\omega_0 (\Omega_+ + \Omega_-)} \qquad;\qquad
{\bf \Lambda_2} = {(\Omega_+ - \Omega_-)^2 - 4\omega_1^2 \over 2\omega_0 (\Omega_+ - \Omega_-)}
\end{eqnarray}
With these definitions the following expressions hold:
\begin{eqnarray}
{\bf \Lambda_1} + {\bf \Lambda_2} = - {\Omega_+ (\omega_1 - \omega_2)\over \omega_0 \omega_2} \quad;\quad
{\bf \Lambda_1} - {\bf \Lambda_2} = + {\Omega_- (\omega_1 + \omega_2)\over \omega_0 \omega_2} 
\end{eqnarray}
\begin{eqnarray}
{\bf \Lambda_1} \Omega_1 + {\bf \Lambda_2}\Omega_2  &=&  + {1\over \omega_0}
\{\omega_0^2 - (\omega_1 + \omega_2)(\omega_1 - \omega_2)\} \\
{\bf \Lambda_1} \Omega_1 - {\bf \Lambda_2}\Omega_2 &=& {\Omega_+\Omega_- \over \omega_0} \\
 {\bf \Lambda_1} \Omega_2 + {\bf \Lambda_2}\Omega_1 &=& - {1\over \omega_0}
\{\omega_0^2 + (\omega_1 + \omega_2)(\omega_1 - \omega_2)\} \\
{\bf \Lambda_1} \Omega_2 - {\bf \Lambda_2}\Omega_1 &=& \{{\omega_1\over \omega_2}\} {\Omega_+\Omega_- \over \omega_0}
\end{eqnarray}
\vskip 0.4cm
\noindent $\bullet$ {\bf Acknowledgment.} This research has been carried out with the partial support of  
{\bf MINECO} under the project MAT2013-46308-C2-1-R and JCyL Project Number SA226U13

\end {document}